\documentclass[letterpaper, 10 pt, conference]{ieeeconf}  

\IEEEoverridecommandlockouts                              

\usepackage{epsfig} 

\title{\LARGE \bf
A Novel Procrustes Analysis Method to Quantify Multi-Joint Coordination of the Upper Extremity after Stroke
}

\author{Khadija F. Zaidi$^{1}$ and Michelle Harris-Love$^{2}$
\thanks{*This work was supported by a Dissertation Completion Grant awarded by George Mason University}
\thanks{$^{1}$Khadija F. Zaidi is a Doctoral Candidate in the Computational Biomedical Engineering concentration of the Department of Bioengineering
        Volgenau School of Engineering, George Mason University, Fairfax, Virginia, USA
        {\tt\small szaidi8@gmu.edu}}%
\thanks{$^{2}$Michelle Harris-Love is with Anschutz Medical Campus, University of Colorado,
        Aurora, Colorado, USA
        }%
}

\begin{document}

\maketitle
\thispagestyle{empty}
\pagestyle{empty}

\begin{abstract}

Upper extremity motor impairment affects about 80\% of persons after strokes. For stroke rehabilitation, upper limb kinematic assessments have increasingly been used as primary or secondary outcome measures. Studying the upper extremity provides a valuable tool for assessing limb coordination, mal-adaptations, and recovery. There is currently no universal standardized scale for categorizing multi-joint upper extremity movement.
We propose a modified Procrustes statistical shape method as a quantitative analysis that can recognize segments of movement where multiple limb segments are coordinating movement. Generalized Procrustes methods allow data points to be compared across an array simultaneously rather than comparing them in pairs. Rather than rely solely on discrete kinematic values to contrast movement, this method allows evaluation of how movement progresses. 
The Procrustes analysis of able-bodied movement showed that the hand and forearm segments moved in a more coordinated manner during initiation. The shoulder and elbow become more coordinated during movement completion. In impaired movement, this coordination between the hand and forearm is disrupted. Potentially mal-adaptive compensation occurs between the upper arm and forearm after movement enters the deceleration phase. 
The utilization of Procrustes analysis may be a step towards developing a comprehensive and universal quantitative tool that does not require changes to existing treatments or increase patient burden.
\newline

\indent \textit{Clinical relevance}— This modified Procrustes Shape Analysis method can be applied by clinicians to motion capture data from patients suffering upper extremity movement deficits to objectively identify multi-joint coordination.
\end{abstract}

\section{INTRODUCTION}

The human upper limb can move around seven degrees of freedom (DOF) and demonstrates motor abundance, i.e. the ability to achieve a task in many different ways. When the human upper extremity reaches towards a target at arm’s length or beyond, the shoulder, elbow, and wrist joints are all involved in coordinating the movement \cite{brown1991}. The neuromuscular control of reaching is computationally complicated; however behavioral or kinematic analyses of the upper extremity can be valuable tools for clinicians in evaluating movement in individuals with neurological disorders (e.g. Stroke) \cite{schwarz2022, ohberg2019}.

Upper extremity motor impairment affects about 80\% of persons after strokes and many activities of daily living. Therefore, being able to use the arm to complete functional tasks is one of the top priorities for persons after strokes, caregivers, and health care professionals. For stroke rehabilitation, upper limb kinematic assessments have increasingly been used as primary or secondary outcome measures. Clinicians may measure quality of movement through goniometry, intertial movement units (IMU), ultrasound imaging, and a variety of kinematic metrics. After a functional impairment, spasticity, restricted range of motion, abnormal muscle synergies, and weakness can all affect reaching \cite{takatoku2014,chiovetto2010}. According to kinematic and kinetic descriptions, arm reaching paths in the paretic arm become less smooth, and elbow and shoulder rotations become less coordinated \cite{mccrea2002}. 

Studying the upper extremity also provides a valuable tool for assessing recovery or retention of rehabilitative effects after the development of a functional limitation. The ability to use the upper limb during functional tasks may improve through compensatory strategies, in which individuals adapt to complete a task in an entirely different way. The ability to distinguish between mal-adaptive movements and compensation would help identify interventions that can influence true recovery \cite{saes2022}. The current standards to indicate coordinated movement include electromyography (EMG), clinical observation, manual muscle testing, and screening tests. The tests most used by clinicians score the upper extremity on the basis of achieving or not being able to achieve certain tasks. These assessments do not yet provide enough information about the strategies and mechanisms that underpin atypical reaching, and are unable to distinguish between true recovery and compensation.

Motion capture analysis has already been integrated into clinical practice as a gold standard for kinematic analysis and is increasingly used clinically to assess quality of movement following injury \cite{montoya2022}. Current biomechanical analysis methods include motion capture systems and accelerometers; however, the use of such strategies is still not standard across clinical practice \cite{collins2018}. While there are many kinematic metrics that can assess quality of movement such as velocity profiles, movement units, smoothness, path error, target accuracy, there is currently no universal standardized scale for categorizing multi-joint upper extremity movement, and current assessments provide little information on how to tailor rehabilitation strategies to an individual patient. Such models already exist for gait analysis and have been proven to be beneficial. 

In this paper we explore the Procrustes Analysis method and how it may be applied to upper extremity kinematic analysis to indicate coordinated movement between multiple limb segments. We aimed to:

\begin{itemize}

\item Apply the Procrustes method as a quantitative analysis that can recognize segments of movement where multiple limb segments are moving in a coordinated fashion,
\item Demonstrate the potential to compare data from able-bodied adults and persons who have developed functional limitations after stroke.

\end{itemize}

A quantitative tool that can be used across clinical practice without having to alter current measurement modalities, convert large amounts of data, or increase patient burden, presents a step towards developing a comprehensive and universal upper extremity analysis.

\section{PROCRUSTES STATISTICAL SHAPE ANALYSIS}

This study describes a novel application of Procrustes Shape Analysis to multi-joint upper extremity kinematic data which can be used across any data-set produced by clinical motion capture systems. Amusingly named after a Greek myth of an innkeeper that would stretch or amputate the limbs of his hapless victims, the Procrustes Method is a multivariate statistical technique of comparing matrices of data points. Procrustes has been utilized since the 1970s in psycho-metrics and has become increasingly popular across disciplines. Procrustes techniques are used in a variety of food research areas, including sensory and consumer evaluations, by dividing outcomes into matrices that can be compared to one another element by element \cite{gower}. 

Generalized Procrustes methods allow data points to be compared across an array simultaneously rather than comparing them in pairs, which is mathematically equivalent to fitting data to a group average. When applied to upper extremity position data, Procrustes can be used to compare each position across a reaching movement made by a persons after stroke to a neural intact control. Current kinematic analyses compare movements by differentiating between discrete metrics such as peak velocity, movement time, target error, etc. The advantage of applying Procrustes analysis is that the overall movement trajectory can be compared point-by-point. Procrustes also presents the potential to monitor recovery by comparing trajectories of a patient over time to assess if movements unfold similarly.

\section{METHODS}

\subsection{Participants}

Unimpaired adult participants $(n=2)$ and persons after stroke $(n=2)$ were recruited to complete reach-to-target tasks while wearing clusters of motion capture markers centered on the hand, forearm, and upper arm. All subjects completed written consent paperwork as approved by the Institutional Review Board of the National Rehabilitation Hospital in Washington, DC. All identifying information was omitted from subject files. Subjects were excluded in the instance of any other neurological disorders or if on medication to prevent seizures. Inclusion criteria included the ability to complete a reaching task and provide informed consent. Persons with stroke were classified as exhibiting either mild or severe impairment based on an Upper Extremity Fugel-Meyer (UEFM) score, where a score of 66 indicates all test tasks can be performed by the subject. Subject information is shown in Tbl. \ref{Demo}. Able-bodied participants and persons with stroke were all right-dominant.

\begin{table}[ht]
\caption{Demographics of Subjects with Stroke}
\label{Demo}
\begin{center}
\begin{tabular}{|c|c|c|c|c|c|}
\hline
Sub & Months Since Stroke  & UEFM & Paretic & Severity\\
\hline
1	&	31	&	31	&	Right	& 	Mild	\\
2	&	51	&	11	&	Right	& 	Severe	\\
\hline
\end{tabular}
\end{center}
\end{table}

   \begin{figure}[b]
      \centering
      \framebox{\parbox{3in}{\includegraphics[scale=0.11]{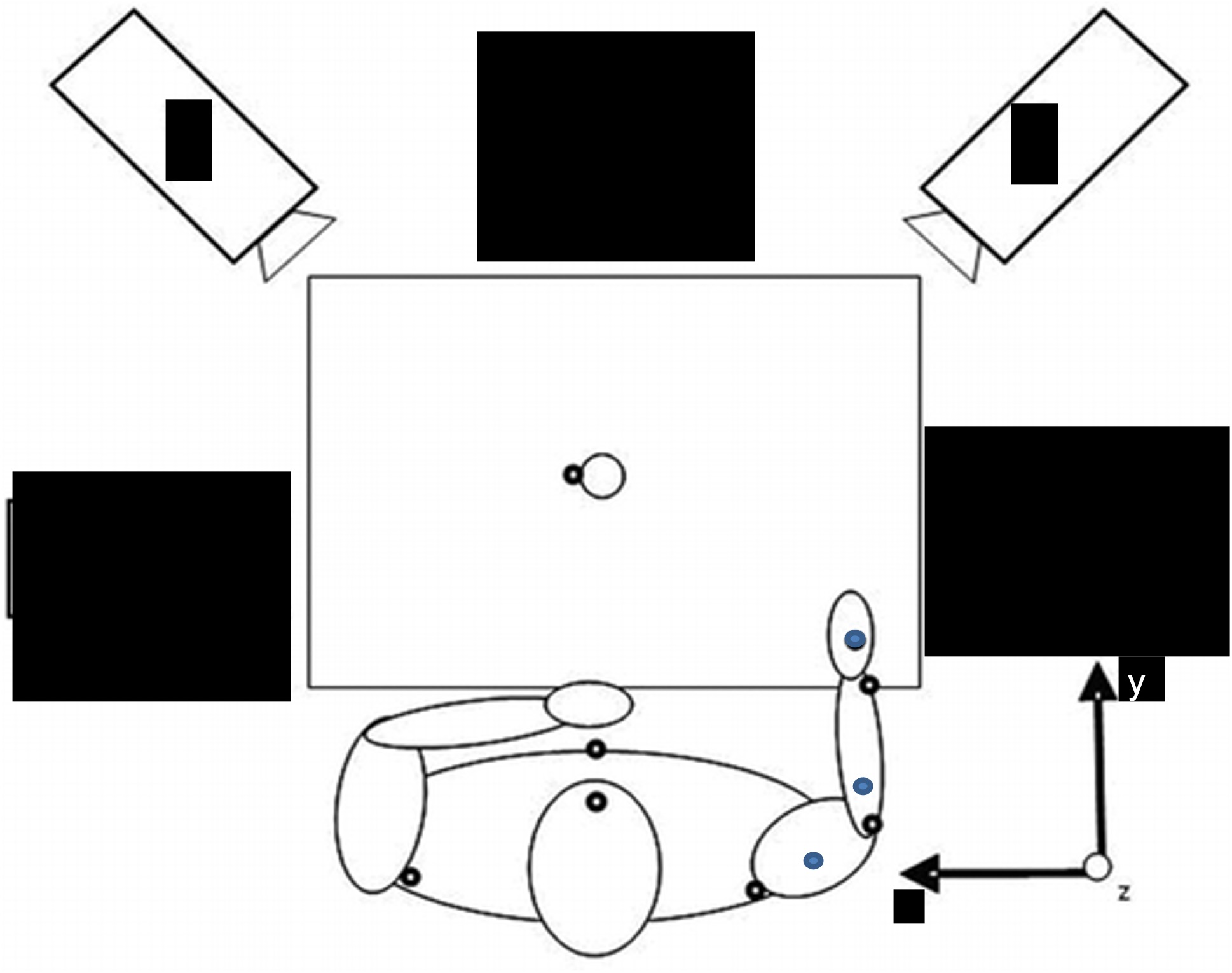}
}}
      \caption{Subject seated at reaching workspace with target placed medially at 80 \% max reach capacity}
      \label{Table}
   \end{figure}

A planar reach-to-target task was selected as a movement primitive that is a sub-movement of many upper extremity functional movements. Each subject completed two blocks of ten reaching movements using both paretic and non-paretic arms for trajectory analysis. All individuals wore trunk restraints and targets were placed well within arm's reach to reduce trunk involvement and shoulder elevation. Targets were placed at 80\% maximal reach distance for all twenty reaches to normalize with respect to the individual's reach capacity. The work space for reaching is depicted in Fig. \ref{Table}. Subjects were instructed to reach for the target at a comfortable pace in response to a "go" signal displayed in front of the subject. Curves for each arm were averaged to produce a representative hand path trajectory. Curves were matched to control subjects using the same arm for movements as each subject’s paretic arm. A fourth order Butterworth filter was applied to the data collected at a sample frequency of 300 Hz, in order to remain maximally sensitive to the desired cutoff frequency of 50 Hz. Studying the standard deviations in orientation angles over the course of movement allows a quantification of the interpatient variability over the course of ten reaching trials per subject. 

\subsection{Data Collection}

\begin{figure}[t]
      \centering
      \framebox{\parbox{2.1in}{\includegraphics[scale=0.11]{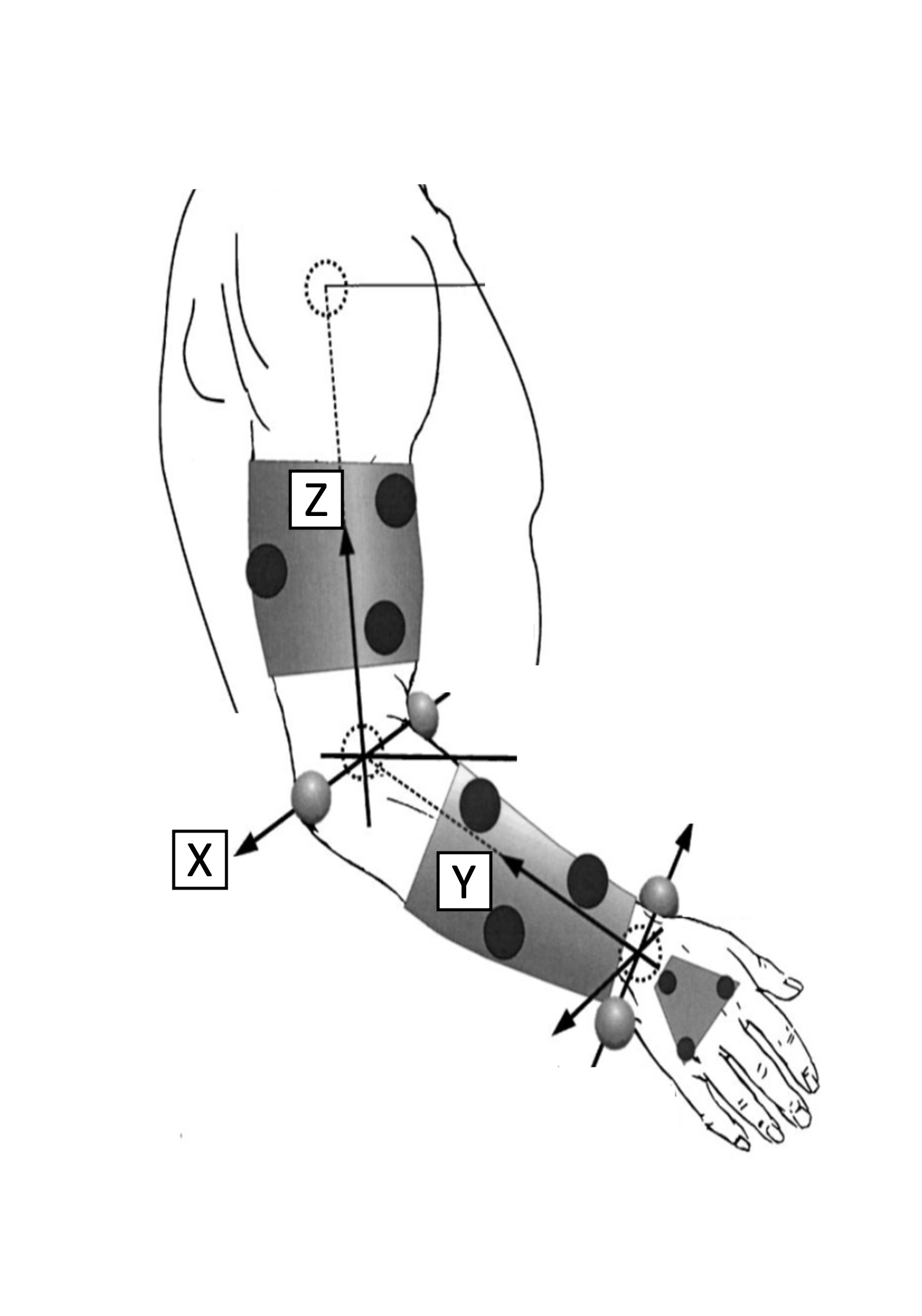}
}}
      \caption{Clusters of three Infrared Emitting Diode markers centered at subject's hand, forearm, and upper arm segments}
      \label{Arm}
   \end{figure}

In order to collect trajectory data for the three segments of the arm, an Optotrak motion capture system was used along with small Infrared Emitting Diode (IRED) marker clusters to represent the upper arm, forearm, and hand [NDI Measurement Sciences] as seen in Fig. \ref{Arm}. Six degree of freedom Optotrak cameras were mounted on the wall and the workspace in front of the subject was digitized as the x-y plane. 7 mm diameter IRED markers were placed as three-marker clusters. The current literature agrees on the placement of three markers in each cluster in order to control for tissue stretch artifacts and to accurately extrapolate the position of the arm. MATLAB was used to filter reaching trajectory data collected from the above described marker clusters in order to represent a clear displacement from the starting edge of the workspace to the target and identify significant landmarks in the hand path trajectory.

\subsection{Sliding Procrustes}

To illustrate the output of the Procrustes method, consider two triangles in two-dimensional space. The data-sets for each triangle consist of the three points that construct each triangle. In a traditional Procrustes analysis, one triangle is used as the reference configuration, while the second triangle is scaled, translated, reflected, and/or rotated as necessary to achieve a similar size and common center. When a data set is compared against a control curve, a linear transformation matrix C, an orthogonal rotation and reflection matrix T, a scaling factor b, and a dissimilarity index D are delivered that would conform that data set to the control.

The able-bodied control trajectories were compared in part and in full to trajectories with motor impairments in order to establish an index of dissimilarity. In this study, the curves were not scaled, since capacity to reach is specific to each subject. A novel modification was made to the Procrustes method by analyzing subsets of the three-dimensional reaching path data in increments of 35 consecutive data points, henceforth referred to as Sliding Procrustes. By comparing reach paths in shorter sub-phases, movement that is kinematically congruent between impaired and able-bodied movement can be highlighted. The index of dissimilarity D, the sum of the squared Procrustes distance between each corresponding element in both curves, represents how dissimilar the two curves may be, and is normalized such that it produces a value between 0 to 1, where 0 represents congruence between curves and 1 represents complete dissimilarity. 

In order to identify limb coherence, a dissimilarity index was produced comparing whole curves of each limb segment, i.e. the hand, the forearm, and the upper arm. 

\section{RESULTS}

   \begin{figure*}[t]
      \centering
      \framebox{\parbox{6.9in}{\includegraphics[scale=0.39]{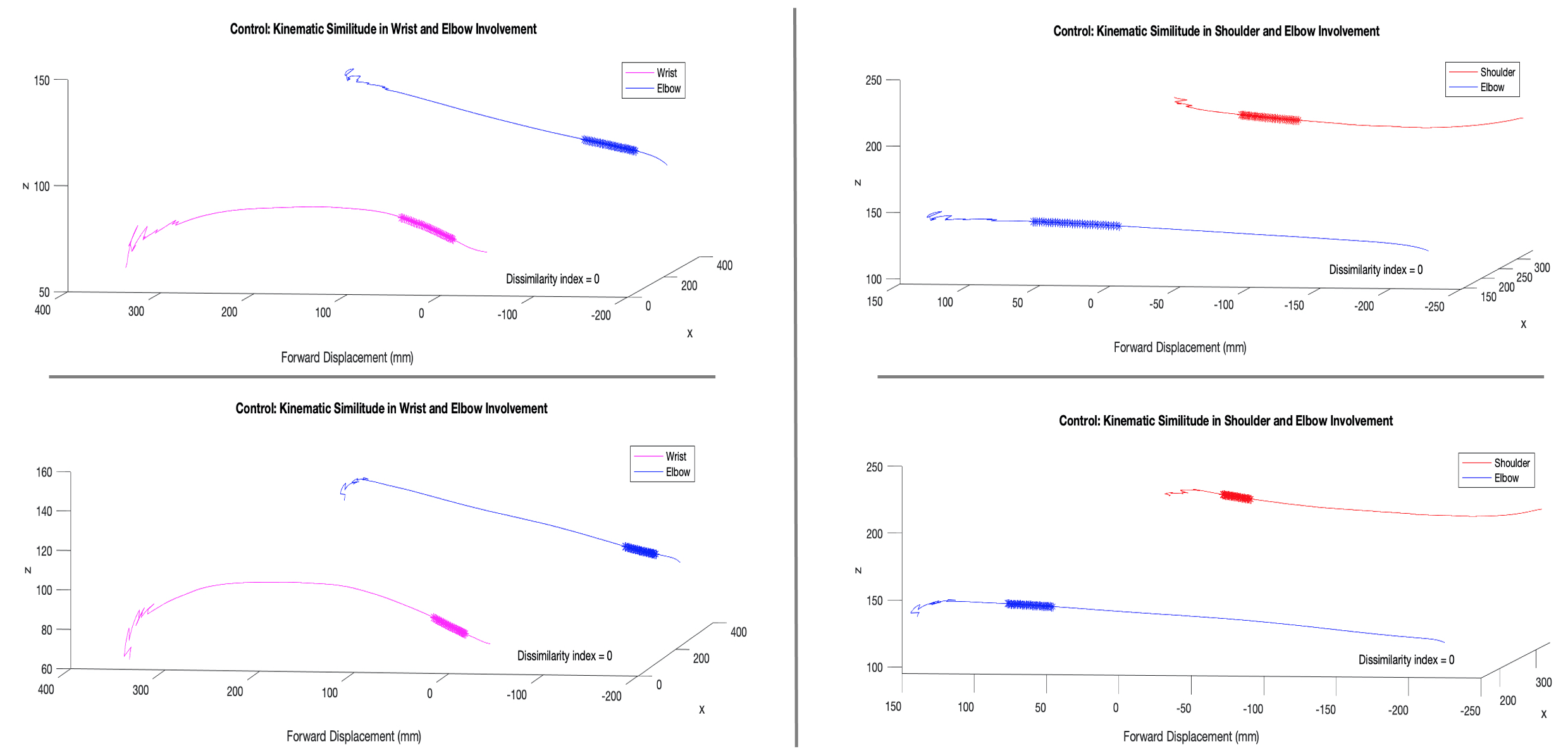}
}}
      \caption{Procrustes comparison of dominant arm of able-bodied subjects demonstrates the upper arm and forearm contributing to a coordinated movement after peak velocity. The hand and forearm appear to coordinate during movement initiation.}
      \label{Controls}
   \end{figure*}

Fig. \ref{Controls} shows the movement completed by the two able-bodied control subjects. The Sliding Procrustes method showed that the hand and forearm segments moved in a more coordinated manner during movement initiation. The shoulder and elbow become more coordinated in the latter portion of movement before the target is reached. Kinematically, these sections of movement are congruent, indicating the limb segments move as one at these time points to achieve motor goals.

\begin{figure*}[t]
      \centering
      \framebox{\parbox{6.9in}{\includegraphics[scale=0.38]{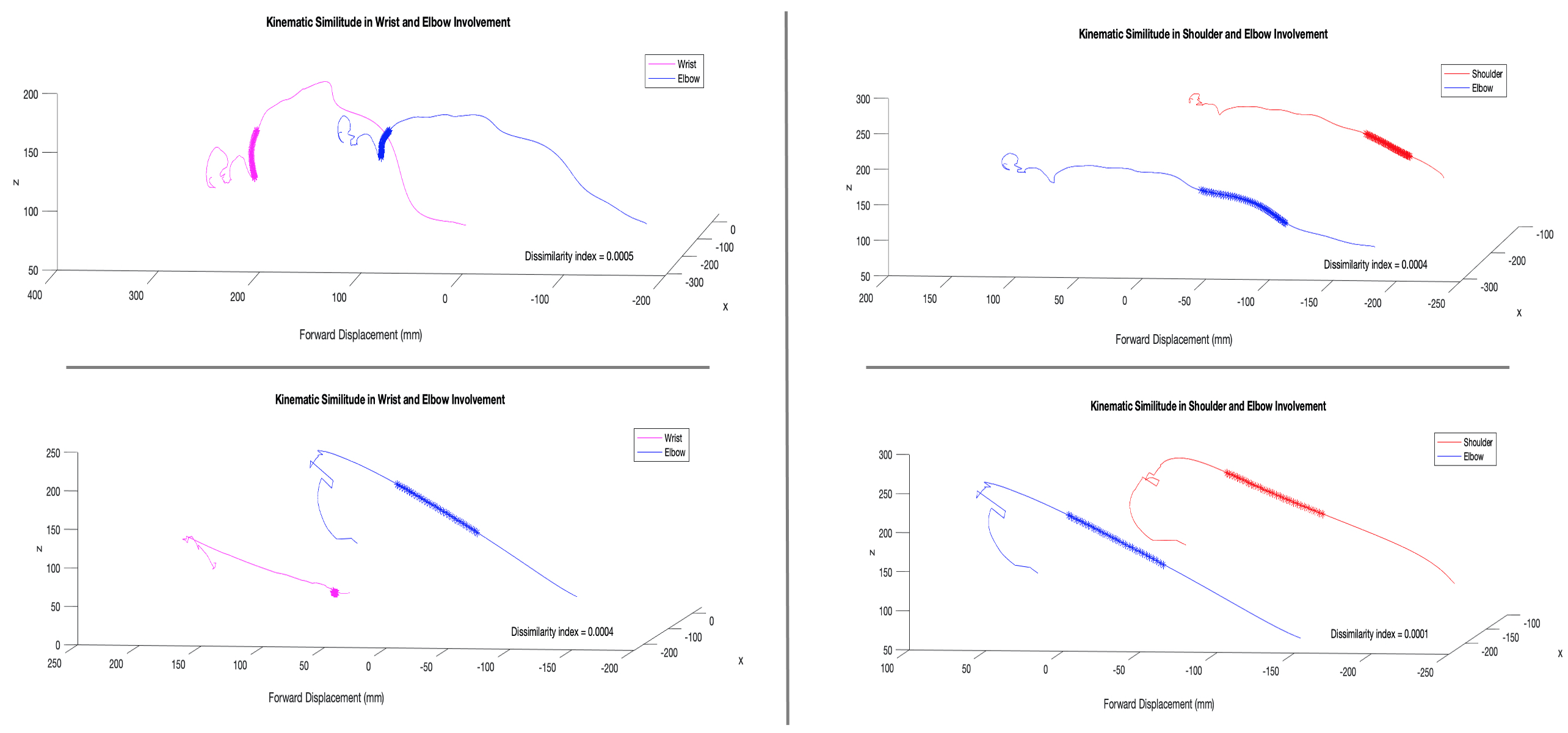}
}}
      \caption{The paretic arm of the subject suffering severe impairment demonstrates potentially mal-adaptive shoulder and elbow involvement early in the movement.}
      \label{Subjects}
   \end{figure*}

Fig. \ref{Subjects} illustrates the output of the Sliding Procrustes analysis when applied to the movement produced by persons with mild and severe impairment. The mild impairment graphs indicate movement of the hand and forearm is similar during movement completion, while the upper arm and forearm are similar earlier in the movement. The severe impairment graphs in the bottom row show no meaningful similarity in the hand and forearm movements. A potential explanation for a lack of coordination may be due to the elbow joint experiencing a diminished range of motion. Patients with more severe distal impairments may incorporate more shoulder or trunk involvement in order to achieve movements in the presence of muscle stiffness. 

Position data from the wrist, or the endpoint, was used to extrapolate peak velocities (m/s), the time between movement initiation and peak velocity (ms), and the time between peak velocity and movement completion (ms). These kinematic measurements are shown in Tbl. \ref{peakvelocity}.

\begin{table}[ht]
\caption{Peak Velocity and Movement Units for Subjects with Stroke}
\label{peakvelocity}
\begin{center}
\begin{tabular}{|c|c|c|c|c|}
\hline
Sub & PV (m/s)  & Time to PV (ms) & Time after PV (ms)\\
\hline
1	&	$0.42 \pm 0.07$	&	$450.48 \pm 119.17$	&	$1143.60 \pm 544.97$ 	\\
2	&	$0.43 \pm 0.03$	&	$214.29 \pm 96.80$	&	$ 736.45 \pm 161.19$	\\
\hline
\end{tabular}
\end{center}
\end{table}

After the reach paths were compared through the Sliding Procrustes method, the temporal locations of lowest dissimilarity were highlighted graphically. Fig. \ref{Controls} shows a complete similarity of movement between the upper arm and forearm, and the forearm and hand, towards the completion of movement and after the peak velocity is achieved. As the target for all subjects was placed at 80 \% of maximum reach capacity, synergistic movement of the elbow and shoulder can be present during precise movements or error correction, which are known to occur in the latter part of movement.

The severe movement case showed the sub-movements that were most similar between limb segments occurred for longer duration and earlier during the movement. The mild movement case showed coordinated movement for shorter durations and towards the completion of movement. The time points of discrete kinematic metrics can be compared against the segments where Procrustes shows trajectories with low dissimilarity. Peak velocity is achieved by the mild impairment subject during the upper arm and forearm co-movement. In the severe impairment case, the peak velocity is achieved before the upper arm and forearm begin moving similarly. The dissimilarity indices were calculated between the upper arm and forearm, the forearm and hand, and the upper arm and hand. An index of 1 represents the highest comparative dissimilarity. Dissimilarity ratios for the dominant limb of a control subject, as well as ratios for the mild and severe impairment subjects are contained in Tbl. \ref{dissimilarity}. 

\begin{table}[ht]
\caption{Dissimilarity Indices of Whole Curve of Upper Arm (UA), Forearm (FA), and Hand (H)}
\label{dissimilarity}
\begin{center}
\begin{tabular}{|c|c|c|c|}
\hline
 & UA/FA Index  & FA/H Index & UA/H Index\\
\hline
Control 1 & 0.0013 & 0.0022 & 0.0057 \\
Control 2 & 0.0012 & 0.0007 & 0.0023 \\
Subject 1 & 0.0039 & 0.0174 & 0.0336 \\
Subject 2 & 0.0073 & 0.1761 & 0.215 \\

\hline
\end{tabular}
\end{center}
\end{table}

The two control subjects demonstrate the greatest dissimilarity in movement occurs between the upper arm and hand during reach-to-target. In contrast, any potential evidence of coordinated movement is between the upper arm and forearm, or the forearm and the hand. The mild impairment subject shows an increased dissimilarity in how limb segments modulate movement, with impairment affecting coordination between the elbow and shoulder. 

\section{DISCUSSION}

Though kinetic measures of joint ranges of motion and coordination have been identified as more effective descriptors of impairment than spasticity scores, there is currently no single method of organizing functional movement into a structured objective framework. The Sliding Procrustes method provides a standardized analysis that can be applied to any clinically produced set of endpoint position data. In the preliminary data, this method identifies the time points at which multiple segments of the upper extremity display congruous movement in relation to when peak velocity is achieved. The dissimilarity ratios support the clinical findings of inter-joint coordination becoming disrupted when movement is impaired. Using the cluster marker set to identify the orientation and movement of the upper arm, forearm, and the hand, the Sliding Procrustes approach can be applied to individual segments of the upper extremity. When the Sliding Procrustes analysis is applied to individual limb segments, it is seen in the control case that the hand and forearm assume the same reaching progression earlier during the movement, while the upper arm and forearm become similar later during the movement. 

Rather than decompose movement into the impulse control, sub-corrections, and limb target control phases, the Sliding Procrustes method was used to compare every subject curve segment of 35 time-points against every segment of the control subjects. The MATLAB script was custom-written to compare every segment and sort by the output dissimilarity index, to find the segments between control and impaired movement that were most congruous. In the mild impairment movement curve, the initiation behavior of movement is extremely similar to that of the control curve. This may imply spinal inputs being responsible for the initiation of a strongly pre-learned behavior, and for these inputs to remain preserved despite stroke pathology. Dissimilarity in such a case would result from difficulty during online error regulation. This supports the theory of initiation and pre-planning control mechanisms either being preserved or possibly recovered post-stroke. In the severe impairment movement curve, similarity does not occur at the initiation of movement. As expected, the deceleration of movement, when the shoulder and elbow joints are known to coordinate to complete the reaching task, is also not similar to neurally intact movement. Due to severity of impairment, it was expected there would be no significantly similar partial curve that would match the control case. As shown by this preliminary data, the Sliding Procrustes approach is limited in its ability to explain how a particular limb segment contributes to the similarity of partial curves, whether through reduced range of motion or mal-adaptive compensation. 

This method may allow for further examination of which segments may be modulated through a synergy; i.e. exhibit similar partial curves at the same time increments. This approach represents an overlap between the trajectory analysis and joint analysis, as it is neither a purely kinematic nor dynamic metric.  It is hoped with the inclusion of more subjects, it will be further examined how and when the individual limb segments coordinate for optimized movement to emerge. Future studies might couple the Sliding Procrustes output with simultaneous electromyography (EMG) of the agonist and antagonist muscles to validate whether the congruent sub-movements stem from synergistic and coordinated movement.

\addtolength{\textheight}{-12cm}   




\section*{ACKNOWLEDGMENT}

K.F. Zaidi thanks all that participated in this study. Thanks are due specifically to Dr. Qi Wei, Dr. Quentin Sanders, Dr. Rachael Harrington, and the MedStar National Rehabilitation Hospital research department.

\end{document}